# OPTICAL PHENOMENA IN MESOSCALE DIELECTRIC THREE- AND TWO-DIMENSIONAL STRUCTURES

I.V. MININ[1], O.V. MININ[1]

[1]*Tomsk polytechnic university*

During the last decade, new unusual physical phenomena in studies on the optics of dielectric mesoscale particles of arbitrary three-dimensional shape and gratings with Mie size parameter q ~ 10 have been discovered. The report provides a brief overview of these phenomena from optics to terahertz, plasmonic and acoustic. We also discuss different particle configurations (isolated or interacting, Janus) and draw a brief overview of the possible applications of such mesoscale structures, in connection with field enhancement, nanoparticles manipulation and super resolution imaging. The presence of a number of interesting applications indicates that a new promising direction has emerged in optics, terahertz, plasmonics and acoustics.

И.В. МИНИН[1], О.В. МИНИН[1]

[1]*Томский политехнический университет*

## ОПТИЧЕСКИЕ ЯВЛЕНИЯ В МЕЗОРАЗМЕРНЫХ ДИЭЛЕКТРИЧЕСКИХ ТРЕХ- И ДВУМЕРНЫХ СТРУКТУРАХ

В течение последнего десятилетия были обнаружены новые необычные физические явления в оптических исследованиях диэлектрических мезомасштабных частиц произвольной трехмерной формы и решеток с параметром размера Ми q ~ 10. В работе дается краткий обзор этих явлений в оптике, терагерцовом, акустическом диапазонах и плазмонике. Мы также обсуждаем различные конфигурации частиц (изолированные или взаимодействующие, Янус) и делаем краткий обзор возможных применений таких мезомасштабных структур для усиления поля, манипуляциями с наночастицами и сверхвысокого разрешения. Наличие ряда интересных приложений свидетельствует о появлении нового перспективного направления в оптике, терагерцовои, акустическом диапазонах и плазмонике.

Точное решение уравнений Максвелла для рассеяния плоской волны на однородной диэлектрической сфере описывается уравнениями Ми [1], включающим параметр размера $q=2\pi R/\lambda$, где *R* - радиус сферы, *λ* - длина волны излучения и зависит от *n* - показателя преломления частицы. Структуры поля с параметром размера $q\sim 10$, лежащие в области между нано- ($q\sim 1$), волновой и геометрической оптикой (q>100), ранее были малоизучены. Однако за последние десять лет в исследованиях по оптике диэлектрических мезоразмерных частиц произвольной трехмерной формы и решеток с параметром $q\sim 10$ были обнаружены новые необычные явления.

К ним относятся:

**эффекты, связанные с возбуждением анапольных мод**

Термин «анаполь» (что по-гречески означает «без полюсов») был введен в физику элементарных частиц Ю.Б. Зельдовичем [2]. Анаполь – это конфигурация двух моментов, тороидального и электрического дипольного моментов, при которой они колеблются в противофазе друг относительно друга. Анапольное состояние проявляется в деструктивной интерференции мультипольного отклика диэлектрической частицы как в дальней, так и в ближней зоне, соответствующее пространственному разделению электрической и магнитной компоненте поля. Поскольку диаграммы направленности тороидального и электрического дипольного моментов практически идентичны, волны, излучаемые ими в такой конфигурации, взаимно гасят друг друга. В то же время, электромагнитная энергия локализуется в ближней зоне такой структуры.

По-видимому, впервые возможность возбуждения анапольного состояния в метаматериалах была показана в работе [3]. При рассеянии света на одиночной сферической частице с высоким показателем преломления можно наблюдать два типа мод анаполей - электрическую и магнитную. Существенные различиями между ними состоят в том, что, во-первых, возбуждение магнитной тороидальной моды может быть реализовано только для больших размерных параметров частицы, которые обычно дают больший вклад, чем остальные возбужденные моды. Во-вторых, эти два типа анаполей имеют существенные различия в пространственной концентрации энергии - электрический анаполь концентрирует энергию внутри частицы, а магнитный анаполь - вне ее [4]. При этом оба типа анапольных мод могут возбуждаться одновременно, что приводит к образованию гибридных анапольных мод [5]. Однако для сферических частиц этому эффекту препятствуют частичные вклады мод более высокого порядка, которые приводят к образованию нетривиальных конфигураций поля с вихрями и соответствующими особенностями внутри и вне частицы. Более того, [6] возможен режим, когда рассеяние малой (q<<1) сферической диэлектрической частицы существенно ниже значения, которое следует из предела Рэлея [7]. В тоже время, анаполь может быть сгенерирован с помощью диполя в присутствии малой (q<<1) сферической частицы с высоким показателем преломления [8].

Возбуждение анапольной моды в диэлектрических частицах имеет практическое применение в нанофотонике, фотонике «на кристалле» для создания новых резонаторов, сенсоров в различных диапазонах частот [9-11].

**оптические нановихри, оптические водовороты, оптические сердца**

Аномальное рассеяние света обладает множеством интересных и необычных свойств, таких как, например, сложная структура ближнего поля, которая может включать в себя оптические вихри, оптические водовороты и другие особенности в наноразмерной области [12]. Так, термин «оптический водоворот» был введен в работе [13], где было показано, что вблизи плазмонного резонанса металлической частицы образуются оптические вихри нанометрового масштаба, похожие на оптический водоворот. В [14] описаны оптические вихри с характерными размерами сердцевины значительно ниже дифракционного предела для частиц с малым размерным параметром q<<1. Было показано, что в точках, соответствующих симметричному квадрупольному резонансу и прямому и обратному рассеянию, в ближнем поле образуются оптические вихревые структуры.

Другим интересным эффектом, связанным с оптическими вихрями, является возможность трехмерного удержания света в субволновом объеме за пределами диэлектрической частицы с размерным параметром q~10, известным как фотонная струя. Это связано с образованием вихрей в ближнем поле, а именно – с сжатием поля за счет появления оптических вихрей на границе частицы в области образования фотонной струи [15].

В работе [16] на основе теории Лоренца-Ми было обнаружено, что сферические мезоразмерные частицы с определенными параметрами размера могут стимулировать чрезвычайно большую напряженность поля в сингулярностях и образовывать две горячие точки вокруг полюсов частицы. Так, усиления интенсивности поля в горячих точках соответствуют значениям 438 и 514 для параметров размера тефлоновой сферы $q$=22.24159 и $q$=28.64159, соответственно. Была также обнаружена удивительно большая циркуляция трехмерных векторов Пойнтинга в форме сердца (рис.1), которую невозможно предсказать с помощью обычного двумерного анализа.

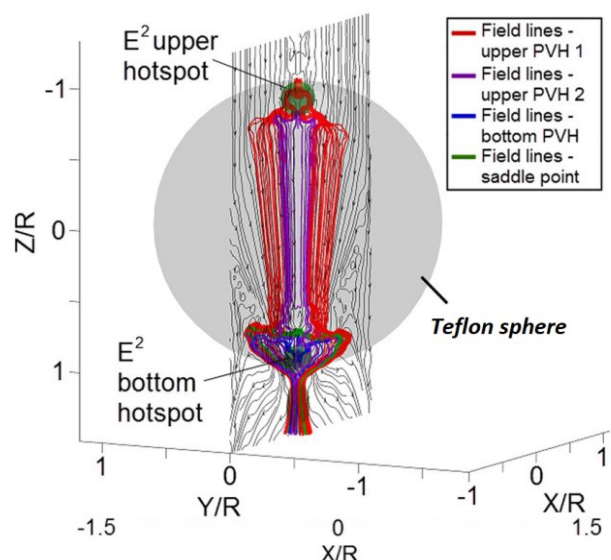

Рис.1. Трехмерный график векторов Пойнтинга, начинающихся в критических точках. Параметр размера частицы *q* = 22.24159. PVH – Горячие точки вектора Пойтинга [16].

Эти новые физические явления позволяют объяснить такие захватывающие эффекты, как формирование фотонной струи и горячих точек.

**Фано резонансы в диэлектрической Ми-резонансной мезофотонике**

Решающую роль в возникновении резонансов Фано играют магнитные дипольные резонансы изолированных диэлектрических частиц. Возбужденная на длине волны магнитного резонанса магнитная дипольная мода диэлектрической частицы может быть более сильной, чем отклик электрического диполя, и тем самым вносить основной вклад в эффективность рассеяния. В [17] продемонстрировано, что слабо рассеивающие мезоразмерные диэлектрические сферы могут поддерживать резонансы Фано высокого порядка, связанные с внутренними модами Ми. Эти резонансы, возникающие при определенных значениях размерного параметра, дают коэффициенты усиления напряженности поля порядка $10^4$–$10^7$. В связи с этими «суперрезонансами» продемонстрировано появление магнитных фотонных струй и гигантских магнитных полей для частиц с высоким (порядка 4 и выше) показателем преломления, которые могут быть привлекательными для многих фотонных приложений.

В [18] был обнаружен эффект сверхусиления фокусировки в сферических частицах с определенными параметрами размера и показателем преломления меньше 2, который обеспечивает примерно в 4000 раз более сильную напряженность поля, чем для падающего излучения и демонстрируют возможность преодоления дифракционного предела, несмотря на высокую чувствительность к величине потерь в материале частицы. Такие сферические частицы имеют уникальное расположение горячих точек на полюсах сферы и обусловлены специфическим поведением внутренних мод Ми. При этом при возрастании потерь в материале сферической частицы, максимальная интенсивность электрического и магнитного полей сравниваются для параметра размера q=22.24159. Для значения параметра q=28.64159 интенсивность магнитного поля примерно в 2 раза выше, чем электрического поля в условиях отсутствия потерь в материале частицы. При этом размер горячих точек меньше дифракционного предела и может быть даже меньше, чем в идеальном случае материала частицы без потерь.

Эффект субдифракционной локализации излучения в двумерной фазовой решетке основан на явлении аномальной аподизации, которому способствуют резонансы Фано, возбуждаемые в мезомасштабной структуре из-за эффективного взаимодействия управляемых резонансов Фабри – Перо и структурных Ми-резонансов [19].

В результате аподизации ступеньки фазовой решетки внутри фазовой ступени образуется открытая резонансная полость с частично отражающими стенками. При определенных параметрах решетки внутри прямоугольной ступеньки возбуждается магнитный мультиполь, составляющая поля которого характеризуется множеством отдельных локальных максимумов, равномерно распределенными вдоль стенок резонатора. В результате вблизи внутренней и внешней поверхностей аподизирующей маски, а также в вершинах ступеньки решетки возникают оптические вихри. Внутри области вихрей входящие оптические потоки направляются и циркулируют в противоположных направлениях вблизи и вокруг внутренних краев фазовой ступеньки. Если для определенного геометрического параметра маски и шага решетки выполняются условия фазового синхронизма встречных оптических волн, то конструктивная интерференция волн приводит к резонансной моде Ми (рис.2).

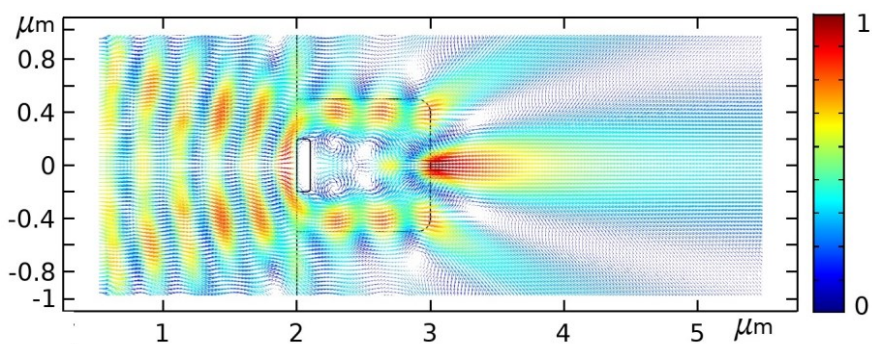

Рис.2. Распределение вектора Пойтинга и оптические вихри в фазовой ступеньке мезоразмерной дифракционной решетке с аподизацией.

Применительно к эффекту Тальбота, описанный эффект позволяет увеличить пространственное разрешение до субдифракционного (около λ/4) и обеспечить высочайший оптический контраст (~22 дБ) [20], что важно для ряда приложений, включая литографию. Позднее похожие подходы независимо начали рассматриваться другими авторами [21].

**фотонные наноструи**

Формально термин «фотонная наноструя» был введен в [22], хотя этот эффект был известен ранее – см., например, [23-26]. Из теории рассеяния света на цилиндрической или сферической частицах известно (см., например, [23,27]), что оптическое поле как внутри, так и вне слабо поглощающей частицы, освещенной световой волной, характеризуется наличием пространственных зон фокусировок, называемых внутренними и внешним фокусами. Их появление обусловлено кривизной поверхности сферической частицы, приводящей к соответствующим деформациям падающего на частицу фазового волнового фронта. Сферическая (или цилиндрическая) частица, таким образом, выполняет роль рефракционной линзы, фокусирующей световое излучение в пределах субволнового объема [28]. При этом, если частица находится в поле сходящейся волны, то область фокусировки смещается в сторону центра частицы и даже может находиться внутри нее [29]. Более того, при облучении частицы структурированными полями, например, вихревым пучком с круговой поляризацией, в формируемой фотонной струе наблюдается обратный поток энергии в фокальной плоскости, зависящей как от топологического заряда оптического вихря, так и от параметра размера сферических частиц [30].

Взаимодействие аберрированных участков волнового поля внутри диэлектрической частицы носит сложный характер и зависит, в частности, от формы частицы, характеристик ее материала, размерного параметра Ми частицы и т. п. При определенном соотношении параметров частицы (оптический контраст материала, размерный параметр частицы) один из внутренних фокусов частицы перемещается к ее внешней границе, формируя вблизи теневой поверхности и вне частицы сильно локализованное поле, названное фотонной струей.

В [22] упоминалось, что «фотонная наноструя … может распространяться на расстояние, превышающее длину волны λ, после выхода из диэлектрического микроцилиндра или микросферы без потерь с диаметром более 2λ [31]. Характерной особенностью фотонной струи является то, что перетяжка пучка может быть меньше классического дифракционного предела [28] - около трети длины волны падающего излучения. Отметим, что такой же размер перетяжки в области фокуса достигается у плоской дифракционной оптики с фокусным расстоянием менее длины волны [32,33]. Однако в общем случае поперечный размер фотонной струи может быть как меньше, так и больше дифракционного предела.

Стоит отметить, что фотонная струя формируется в области пространства, где существенную роль играют эффекты рассеяния в ближней зоне (затухающие, эванесцентные поля). Поэтому, в частности, для частиц с размерами порядка длины волны излучения согласие между приближением геометрической оптики и строгой электромагнитной теории менее точное, поскольку эванесцентное поле не учитывается в геометрической оптике [34]. Обычно эта ближняя область находится на расстояниях, не превышающих нескольких диаметров частиц, и характеризуется заметным вкладом радиальной составляющей оптического поля. В свою очередь, это условие накладывает ограничения на диапазон размеров диэлектрических частиц, который должен быть порядка нескольких длин волн и даже равняться длине волны излучения [15,35], т.е. иметь мезомасштабные размеры. Из этого, в частности, следует, что конкретное значение длины падающей волны не является критическим, пока выполняются условия мезомасштабности.

В [36,37] впервые было показано, что фотонная струя может быть сформирована несферическими и несимметричными диэлектрическими мезомасштабными частицами произвольной формы как в режиме пропускания, так и в режиме отражения. Это позволяет лучше понять свойства фотонных струй и возможные стратегии настройки таких волновых структурированных пучков. С другой стороны, не осесимметричность и несферичность формы частицы приводит к появлению и некоторых новых оптических эффектов, которые также будут кратко рассмотрены ниже.

На сегодня спектр применений эффекта фотонной струи достаточно широк. Для оптических манипуляций этот эффект начал использоваться с начала 2000 годов [38], их можно также использовать для обнаружения и манипулирования биологическими объектами [39], биомедицине [40, 41], микро- и нано- манипуляций [42], оптических ловушках [43,44], Раман спектроскопии [45,46], и многих других.

**Фотонная струя: электрическая или магнитная?**

Для диэлектрических частиц с большими параметрами размера интересно отметить, что фотонная струя является скорее «магнитной», чем «электрической» струей [15]. То есть, когда формируется фотонная струя, усиление магнитного поля может быть намного больше, чем то, которое соответствует электрическому полю. В качестве примера на рис. 3 [15] показано распределение напряженности электрического поля $E^2$ вместе с напряжением магнитного поля $H^2$ и модулем вектора Пойнтинга S для сферы с n~1,5 и параметром размера q~10. Видно, что максимальная интенсивность магнитного поля в области локализации излучения примерно в 2 раза выше, чем интенсивность электрического поля.

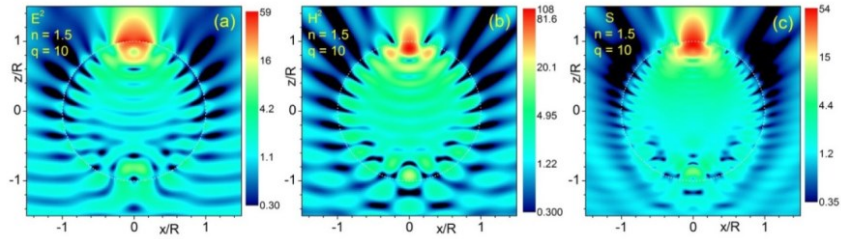

Рис. 3. Распределение интенсивности в плоскости {x-z} (a) электрического поля $\mathbf{E}^2$, (b) магнитного поля $\mathbf{H}^2$ и (c) модуля вектора Пойтинга $\mathbf{S}$ для сферической частицы с показателем преломления $n$ =1.5 и параметром размера $q$ =10 [15].

Обычно электрические дипольные переходы на $10^4 \sim 10^5$ сильнее, чем магнитные дипольные переходы. Однако в отличие от плазмонных частиц, первый резонанс диэлектрических сферических частиц является магнитодипольным резонансом и имеет место, когда длина волны света внутри частицы равна диаметру λ/n~2R. При этом условии поляризация электрического поля антипараллельна на противоположных границах сферы, что приводит к сильной связи с циркуляционными токами смещения, в то время как в середине магнитное поле колеблется вверх и вниз [47].

Следовательно, проявление усилений магнитного поля, ранее обнаруженных для наночастиц с высоким показателем преломления [48,49], возможно и для частиц с относительно малым показателем преломления $n$<2, если их размер больше длины волны (R>λ), что характерно для области существования фотонной струи [15,50]. Это свойство можно использовать, например, для усиления излучения, характеризуемого магнитодипольным переходом [51].

**эффекты генерации гигантских магнитных полей**,

Новый физический эффект - оптический суперрезонанс в мезоразмерных диэлектрических сферах, обусловленный резонансом Фано высокого порядка [18,52], может стать новым способом достижения сверхвысоких магнитных полей. Это обусловлено следующими факторами. Важной особенностью резонансов Фано высокого порядка в таких частицах является высокая степень локализации поля, превышающей дифракционный предел, как внутри частицы, так и на ее поверхности. Последнее связано с образованием областей с высокими значениями локальных волновых векторов, аналогично эффектам суперосцилляции [53,54]. Согласно этой теории, локальный волновой вектор можно понимать как локальный градиент фазы $K_l=\Phi=\nabla E/E$. Высокие значения $K_l$ могут быть созданы, например, в оптике свободного пространства, включая вихри и узлы. Это следует, например, из принципа неопределенности Гейзенберга $\Delta E \Delta t \geq \hbar/2$, который в терминах числа фотонов и их фазы можно переформулировать как $\Delta N \Delta \Phi \geq 1/2$. Дифференцируя эту формулу, можно найти, что $K_l \sim \nabla N/N^2$, т.е. высокие локальные волновые числа могут быть достигнуты в окрестности оптического вихря, представляющего собой сингулярность (точку нулевой интенсивности) с фазой циркулирующего вокруг этой точки поля [52].

В диэлектрических мезоразмерных частицах оптические вихри возникают, когда размерный параметр q превышает некоторое значение, зависящее от его показателя преломления [15]. Соответственно, кольцевые токи в соответствии с законом Био-Савара, создают магнитные поля. Внутри диэлектрической частицы магнитное поле может быть усилено более, чем на 4 порядка, что может дать значения магнитной индукции порядка $10^5$ Тл, а это близко к межатомным магнитным полям [52]. Заметим, что такой уровень магнитного поля сравним с магнитно-кумулятивными генераторами [55,56]. При таких полях можно ожидать эффекты магнитной нелинейной оптики, где изменение показателя преломления вызывается чисто магнитными эффектами. Однако такая магнитная нелинейная оптика может быть реализована при выполнении двух условий: 1) диссипация достаточно мала и 2) магнитный нелинейный отклик значительно превышает электрический нелинейный отклик из-за нелинейности $\varepsilon=\varepsilon(E)$ [52].

Рассмотренные эффекты суперрезонансов Фано высокого порядка являются привлекательной основой для таких перспективных направлений, как, например, усиление эффектов поглощения, абляция, вызванная магнитным давлением и ряд других.

### тераструи

Термин «тераструя» был введен в работе [57] применительно к эффекту фотонной струи в СВЧ и/или терагерцовом диапазоне. Основное отличие от оптики заключается в том, что в ТГц диапазоне существенную роль играет коэффициент поглощения материала частицы, которые может превышать оптический на несколько порядков и в отличие от оптики частица почти всегда не может быть рассмотрена как слабопоглощающая. Отметим, что первое экспериментальное подтверждение эффекта фотонной струи было выполнено в СВЧ диапазоне [58].

Диэлектрические кубические частицы имеют определенные преимущества перед сферическими частицами [35]: герерируемая сферическими частицами фотонная струя более эллиптична, чем у кубических частиц при линейной поляризации; с увеличением размера кубоида и диаметра сферы длина струи увеличивается для кубоидов и уменьшается для сфер; кубоиды позволяют формировать фотонные струи с минимальными размерами частицы от 0,5 длины волны, а сферические - с минимальным диаметром более длины волны.

В работе [57] было продемонстрировано, что при уменьшении показателя преломления трехмерных и двумерных кубоидов, тераструя («теранож» в двумерном случае) перемещается изнутри наружу структуры, обеспечивая субволновой размер перетяжки пучка и высокую интенсивность, до 10 раз превышающую интенсивность освещающего пучка. При этом свойства диэлектрической кубической частицы как аналога плоской линзы с размерами порядка длины волны сохраняются при наклонном падении плоской волны как на основной частоте, так и на кратных гармониках [59].

Более того, аналогично оптическому диапазону [28], была также продемонстрирована возможность усиления обратного рассеяния с увеличением на ~ 1,53 дБ и ~ 10 дБ для металлических частиц диаметром $d_1=0,1\lambda$ и $d_2=0,25\lambda$ соответственно, помещенных в область тераструи [57].

Среди интересных областей применения эффекта тераструи можно отметить повышение чувствительности ТГц детекторов за счет локализации падающего на них излучения в субволновом объеме [60-62], повышение чувствительности открытых СВЧ резонаторов [63], использование мезоразмерных частиц, генерирующих тераструю, в качестве диэлектрической малогабаритной антенны в перспективных системах ТГц связи, в том числе 5G-6G [64,65] и другие.

### аномальный сдвиг фазы Гои

В проблеме получения сверхразрешения дополнительная информация о фазовых полях может помочь снять некоторые ограничения. Фазовый сдвиг Гои [66] был исследован для сферических частиц с параметром размера $q=9.34$ [67] и $q=21.4$ [68]. Было показано, что наблюдаемый фазовый сдвиг вдоль оси распространения фотонной струи представляет собой комбинацию классического сдвига фазы Гои и фазового сдвига, вызванного преломлением на сферической частице.

В тоже время, для кубических частиц с параметром размера q=3 (размер грани куба равен длине волны), формирующих фотонную струю, было обнаружено, что модель фазового сдвига Гои, основанная на фокусированном гауссовом пучке, не может объяснить фазовый сдвиг Гои фотонной струи [69]. Поэтому механизм формирования локализованной области излучения на основе тераструи (фотонной струи) отличается от такового для классической линзы или сферической частицы.

Более того, в [70] было сообщено об обнаружении аномальной асимметрии фазового сдвига Гои в фотонной струе при наклонном падении ТМ-волны на диэлектрический кубик. Эта асимметричная фазовая аномалия Гои вызывает неизвестный ранее угол отклонения между нормалью фазового распределения фазы Гои и направлением распространения падающего пучка в месте, где формируется фотонная струя. Этот угол отклонения экспоненциально уменьшается по мере распространения волн на расстояние в несколько длин волн.

Понимание таких фазовых аномалий позволяет локализовать излучение в пространственных областях, меньших, чем классический дифракционный предел разрешения и может найти применение в микроскопии, оптическом захвате и манипулировании наночастицами, интерферометрии [71].

**эффекты аномальной аподизации**

Известно, что аподизация амплитудной маской центральной части линзы уменьшает поперечный размер пучка в фокусе, однако интенсивность в фокусе падает из-за блокировки части падающей волны [72]. Эта же тенденция сохраняется и для сферических мезоразмерных частиц [73,74].

Эффект аномальной аподизации (уменьшение размера перетяжки фотонной струи при увеличении ее интенсивности) был впервые обнаружен для мезоразмерных частиц различной несферической формы [75]. Явление заключается в том, что введение маски на облучаемой стороне частицы приводит к меньшему количеству оптических вихрей около теневой поверхности этой частицы. Аподизирующая маска также увеличивает эффективную числовую апертуру частицы-линзы, что эквивалентно увеличению показателя преломления материала частицы [76-78].

**Контроль тангенциальной компоненты**

Управлять параметрами фотонной струи возможно за счет контроля тангенциальной составляющей электрического поля вдоль боковой поверхности диэлектрической частицы, например, с помощью металлической маски, размещенной вдоль ее боковых поверхностей [79,80]. При наличии металлических экранов вдоль боковой поверхности диэлектрической частицы возникает отраженная волна, так как соответствующая составляющая поля исчезает в области контакта. В результате отражения амплитуда тангенциальной составляющей электрического поля увеличивается в направлении, противоположном направлению распространения волны, и ослабевает в направлении распространения волны [79]. Эти эффекты приводят к смещению области концентрации плотности потока мощности поперек и в направлении распространения волны [81]. Это позволяет контролировать фокусное расстояние фотонной струи, ее протяженность вдоль направления распространения излучения и уменьшать размер перетяжки в области фокуса. С другой стороны, наличие экрана с одной боковой стороны частицы позволяет формировать фотонные крючки и осуществлять угловое сканирование фотонной струей [80].

**Волноводные системы на основе массива диэлектрических частиц**

Известно, что первый экспериментальный образец линзового волновода с набором идентичных линз был разработан Губо и Собелом еще в 1961 году [82,83]. В оптике эффекты периодической фокусировки света изучались для цепочек слабопоглощающих сфер диаметром более 2 мкм (размерный параметр q>12) и с показателем преломления более 1.3. Было показано, что эти эффекты обусловлены существованием так называемых фотонных мод, индуцированных фотонными струями, с периодом, примерно равным диаметру двух сфер, расположенных вплотную к друг другу [84-86].

В [87] был предложен другой тип волновода, состоящего из цепочки трехмерных кубических мезоразмерных частиц (с размером грани, равным длине волны), разделенных воздушным зазором. Было продемонстрировано существование фотонных периодически фокусированных мод, индуцированных тераструями вдоль цепочки таких частиц. Более того было показано, что предложенная структура имеет надежные характеристики и мало чувствительна к потерям в материале частиц (вплоть до значения тангенса потерь около 0.1, что примерно на 5-6 порядков больше, чем в оптике) благодаря тому, что пространственное разрешение тераструй вблизи теневой поверхности каждого куба не сильно ухудшается при существенном увеличении потерь.

По сравнению с оптическим волноводом на основе сферических частиц, в волноводе на основе кубических частиц отсутствует требование контакта микросфер между собой. Для сферических частиц же контакт может быть реализован с помощью микросварки, но это приводит к ухудшению свойств волновода из-за микротрещин, образующихся за счет оплавления материала в области контакта частиц [88,89].

Эффекты периодической фокусировки света также наблюдались для цепочки мезоразмерных частиц с показателем преломления около 1 [90]. В данном случае реализуется волновод с градиентным показателем преломления, то есть показатель преломления постепенно уменьшается от его центра к краю волновода из-за специфической формы частиц (сфера, куб) при постоянном показателе преломления внутри частиц (рис.4). При этом хотя падающее излучение распространяется с одинаковой скоростью на входе в волновод, время достижения выхода из волновода разное, так как скорость распространения волны на краю волновода больше, чем в центре, что приводит к модальной дисперсии. Фактически, такой тип волновода можно рассматривать как аналог волновода с градиентным (параболическим) (GRIN) законом изменения показателя преломления [91,92].

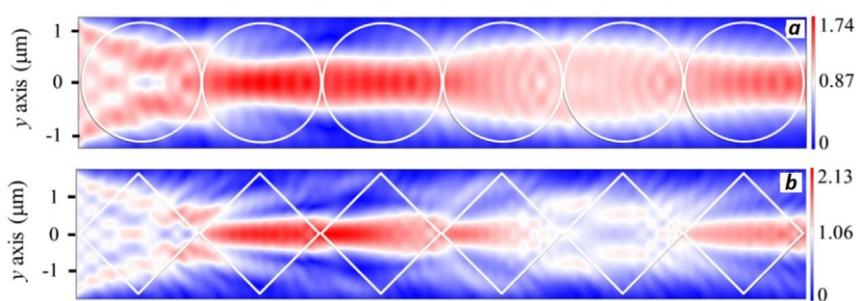

Рис.4. Волновод на основе цепочки (a) сферических (n=0.87) и (b) кубических (n=1.06) частиц.

Другая интересная возможность развития упомянутых приложений связана с использованием мезоразмерных частиц, легированных оптически нелинейными и/или активными материалами, что должно приводить к очень интересным оптическим параметрическим и лазерным эффектам в таких структурах, а также может быть использовано для повышения разрешения изолированной частицы в дальней зоне.

**периодически фокусированные моды, индуцированные Тальбот эффектом**

Эффект Тальбота [93] является одним из основных дифракционных явлений, который возникает для периодических диэлектрических структур при освещении источником когерентных волн. Например, фокусы матрицы сферических микролинз периодически самоизображаются в плоскостях Тальбота, а количество фокусов умножается в дробных плоскостях Тальбота [94]. Однако до недавнего времени было неясно, справедлив ли классический эффект Тальбота для системы из массива мезоразмерных частиц. Учитывая, что диэлектрические частицы, формирующие фотонные струи, обеспечивают более высокий уровень интенсивности в фокусах с размером пятна вплоть до субволнового, была рассмотрена возможность формирования эффекта Тальбота на основе матрицы мезоразмерных частиц. При этом в качестве демонстраций возможности метода, исследования проводились в терагерцовом диапазоне на длине волны λ=0.5 мм с иммерсией частиц диаметром 2 длины волны в диэлектрическую среду с показателем преломления 1.43 и тангенсом потерь 0.002 (рис.5).

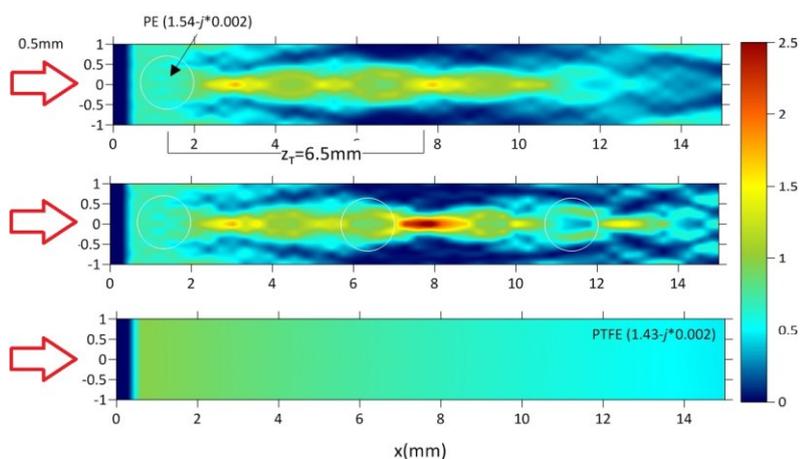

Рис.5. периодически фокусированные моды, индуцированные Тальбот эффектом: дробный Тальбот эффект от решетки сферических частиц, пространственный волновод с размещением сферических частиц в Тальбот-фокусах.

Дробный эффект Тальбота хорошо виден на верхней картинке - фокусы дифракционно - ограниченных пятен повторяются на расстояниях около 2/3 от классического расстояния Тальбота $Z_T=2d^2/\lambda$. Однако интенсивность в фокусах падает с ростом расстояния распространения пучка. Симуляции показывают, что при увеличении показателя преломления частицы наблюдается сдвиг на половину периода в изображении Тальбота. Это, по-видимому, новое явление, о котором ранее не сообщалось в исследованиях этого эффекта на основе матрицы классических сферических линз. Если по аналогии с волноводом, описанном выше, в каждый из фокусов разместить диэлектрическую частицу, то интенсивность излучения в дальних фокусах и контрастность удается повысить по сравнению с матрицей одиночных частиц во входной плоскости. В тоже время, отличие данной концепции состоит в том, что в волноводе на основе линейной матрицы частиц периодичность поля создается фотонной струей, а в данном случае – за счет Тальбот эффекта. Таким образом, концепция пространственного волновода на основе матриц диэлектрических частиц позволяет увеличить расстояние распространения матрицы локализованных пучков в поглощающей среде с разрешением вплоть до превышающим дифракционный предел на основе фотонных периодически фокусированных мод, индуцированных Тальбот эффектом и поддерживаемых фотонными струями вдоль цепочки мезомасштабных трехмерных диэлектрических частиц, разделенных зазорами с материалом среды.

**Фотонная наноструя с зеркальным отражением**

Фотонная наноструя с зеркальным отражением - это особый тип электромагнитной субволновой пространственной локализации излучения в ближнем поле, возникающий в результате конструктивной интерференции прямых и распространяющихся назад волн, сфокусированных прозрачной диэлектрической мезоразмерной частицей, расположенной рядом с плоским отражающим зеркалом и направлена навстречу направлению падения волнового фронта [36,95].

Например, выбором толщины диэлектрической частицы, расположенной на металлическом экране, удается сформировать фотонную струю с эллиптичностью 1,07, т. е. почти осесимметричный фокус при линейной поляризации. Также было показано, что возможно управлять пространственным положением фотонной струи вплоть до ее параллельности плоской поверхности металлического экрана путем поворота всей структуры относительно направления падения излучения [36, 95]. Такой режим формирования структурированных полей возможен для диэлектрических мезоразмерных частиц различной формы, в том числе полусферической [96].

Была также продемонстрирована применимость принципа Бабине дополнительных дифракционных структур для формирования фотонных струй в режиме отражения в ближнем поле [97]. Такие структуры открывают больше возможностей для создания фотонных струй с требуемыми и управляемыми свойствами, такими как фокусное расстояние, ширина, длина, максимальная напряженность поля и эллиптичность струй.

В тоже время, режим формирования фотонных струй в отраженной от преграды волне позволяет обойти ряд ограничений на показатель преломления материала частицы, связанного с областью существования эффекта фотонной струи в проходящем излучении [15]. В [98] было продемонстрировано, что фотонная струя может формироваться в направлении, противоположном направлению распространения падающей волны при использовании частицы с показателем преломления, близким к единице, расположенной на диэлектрической (с высоким показателем преломления) или металлической подложке. При этом может быть достигнута ширина пучка менее дифракционного предела. Анализ распределений вектора Пойтинга показал, что это явление связано многократной циркуляцией потока мощности в частице и ее окрестностях. Следует отметить, что выбором контраста показателя преломления частицы и подложки возможно регулировать параметры области локализации излучения и ее положение одновременно в отраженной и прошедшей волнах. Для примера на рисунке 6 показаны распределения интенсивности электрического поля при падении плоской волны на сферическую частицу с показателем преломления около 1 (n=1.077), расположенной на диэлектрической подложке с показателем преломления n=3.83 и n=4.08 соответственно. Если в первом случае наблюдается фотонная струя в отраженной волне, промодулированная стоячей волной, и слабая область локализации в прошедшей волне, то во втором случае – рефракционная фокусировка в прошедшей волне и слабая локализация в отраженной волне.

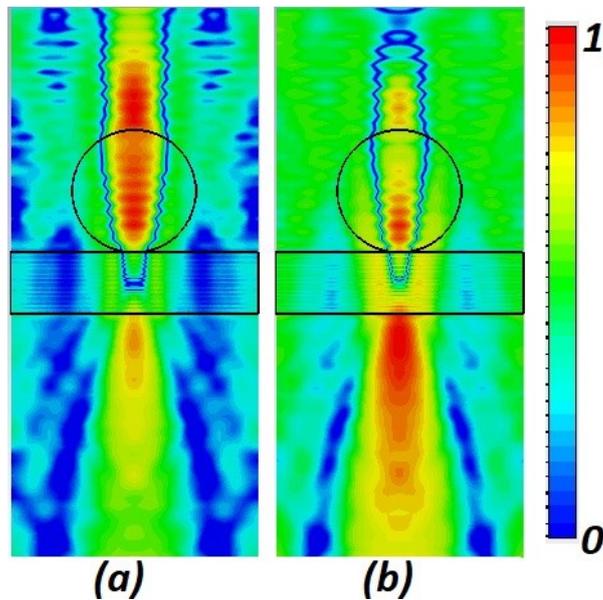

Рис.6. Формирование фотонной струи для сферы с n=1.077, расположенной на диэлектрической подложке: n=3.83 (a) и n=4.08 (b)

Другое уникальное свойство фотонных струй в режиме «на отражение» заключается в ее пространственной субволновой локализации и высокой интенсивности при использовании мезоразмерных частиц с относительно высоким контрастом показателя преломления, когда обычная фотонная наноструя не формируется (n>2) [99]. Фактически, в этой конфигурации осуществляется двойная фокусировка плоской падающей волны с помощью частицы, расположенной вблизи с плоским металлическим зеркалом. Эта возможность преодолеть фундаментальное ограничение максимально допустимого контраста показателя преломления материала частицы обусловлено тем, что при зеркальном отражении излучения от зеркала и последующей фокусировке диэлектрической частицей в ней образуются крупномасштабные оптические вихри с встречно-циркулирующей оптической энергией, вносящее вклад в экстремальную концентрацию оптического поля.

На основе такой фотонной струи была предложена физическая концепция оптической ловушки [99], в которой обеспечивается вдвое более высокая устойчивость к броуновскому движению захваченной наночастицы по сравнению с обычными ловушками на основе фотонных струй [100]. При этом модуляция фотонной струи стоячей волной при необходимости может быть исключена за счет управления интерференционными эффектами [101].

Управление свойствами пространственного распространения узких световых пучков, такими как расходимость, фокусировка, являются основными задачами в оптике и фотонике. Другой интересный эффект формирования фотонных струй в рассматриваемом режиме состоит в том, что тонкая прямоугольная структура диэлектрик-металл может выполнять функцию плоского фокусирующего зеркала [36, 37, 102]. Новый тип планарного устройства без заранее определенной оптической оси [102] позволяет осуществить фокусировку в ближней зоне, где структурированный отраженный пучок демонстрирует бездифракционное распространение на расстоянии длины фотонной струи и может значительно повысить применимость структурированных фотонных систем для управления распространением светового луча в мезомасштабных фотонных схемах. Отметим, что данный тип ближнепольного фокусирующего плоского зеркала существенно проще ранее рассмотренного в [103, 104] плоского фокусирующего зеркала с поперечной инвариантностью и основан на других физических принципах.

**эффекты преодоления дифракционного предела** и **повышения качества изображения**

Возбуждение режима шепчущей галереи в микросфере приводит к увеличению разрешения до ~0.25λ [105]. В [52] было продемонстрировано, что на резонансной частоте, соответствующей параметру размера сферической частицы q=26.94163 может быть достигнут размер фокальной горячей точки около 0.22λx0.37λ для линейной поляризации падающего поля при интенсивности более 43000. При этом достигаемые параметры фокусировки очень чувствительны к параметру размера сферической частицы – изменение размерного параметра q на $10^{-4}$ приводит к уменьшению интенсивности на 4 порядка. Аналогичный эффект резонансного сверхразрешения ранее был рассмотрен в [106]. Интересно отметить, что близкое по размеру фокальное пятно 0.27λx0.6λ для линейной поляризации падающего излучения можно получить в нерезонансном режиме для особой конфигурации зонной пластины [107] с малым уровнем боковых максимумов, но при максимальной интенсивности поля в 1000 раз меньшей (рис.7).

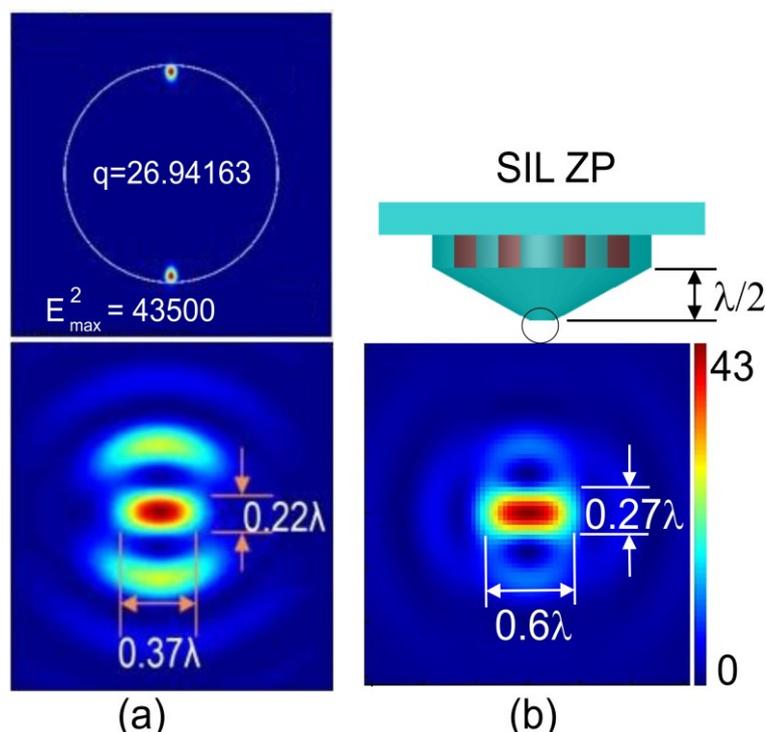

Рис.7. Формирование области локализации излучения для сферической частицы на резонансной частоте и зонной пластины с иммерсионным конусом на нерезонансной частоте при линейной поляризации.

Эффекты генерации горячих точек на полюсах сферических частиц могут быть применены для обнаружения нано-частиц, включая вирусы, когда их нахождение в области горячей точки должно приводить к сдвигу резонансной частоты.

Отметим также, что переход от обычных материалов к метаматериалам позволяет для диэлектрических частиц с размером порядка длины волны добиться суперразрешения - размера перетяжки фотонной струи (тераструи) вплоть до 0.2 длины волны [108], что гораздо меньше классического дифракционного предела. В тоже время достичь эффекта уменьшения размера перетяжки фотонной струи возможно и за счет применения градиентных материалов частицы [109]. Данные эффекты могут помочь решить некоторые аспекты оптической метрологии сверхвысокого разрешения [110].

В [111] обсуждается обнаруженный новый механизм локализации поля в мезоразмерных Янус [112] частицах на основе усеченных сфер или цилиндров. Показано, что электрическое поле на плоской поверхности таких частиц имеет резкие резонансы в зависимости от глубины удаленного сегмента сферы или цилиндра. Эти резонансы связаны с возбужденными волнами шепчущей галереи (МШГ), вызванных усечением части поверхности частицы. Оптимизация этого эффекта для цилиндрических частиц позволяет достичь сверхразрешения на их плоской поверхности.

Физические принципы усеченной сферической линзы, для которой имеет место фокусировка без аберраций, также известная как линза Вейерштрасса, основаны на сжатии формируемого пучка путем уменьшения угла преломления проходящего излучения, измеренного от оптической оси. Это происходит, когда сфера усечена до толщины $d=R*(1 + 1/n)$, где R - радиус сферы, а d соответствует апланатическому фокусу [113,114]. Однако для малых значений усечений (малого размера d) наблюдается перераспределение поля из-за сильного возбуждения МШГ и усиления как электрического, так и магнитного полей. Сингулярность, связанная с разрывом фазы на линии пересечения сферической (или цилиндрической) поверхности с плоской поверхностью, приводит к изменению закона Снеллиуса на обобщенные законы отражения и преломления [115]. По этому закону происходит изменение критических углов полного внутреннего отражения и при некотором значении фазового градиента существует критический угол падения [115]. Этот фазовый градиент зависит от толщины усеченного элемента d и показателя преломления n. На рис. 8 показан результат интерференции двух эванесцентных волн. При этом было обнаружено, что эффективность возбуждения таким новым элегантным методом волн шепчущей галереи сильно зависит от значения усеченного сегмента d.

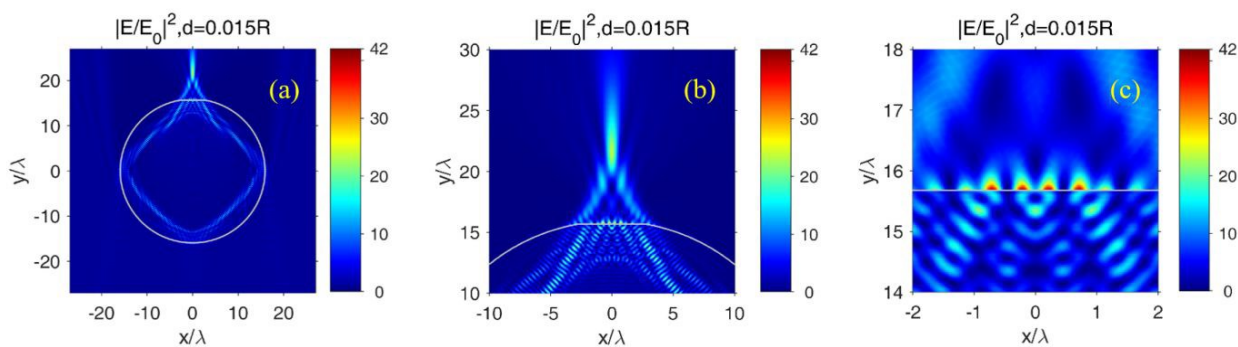

Рис.8. Распределение интенсивности поля для резонансного значения усеченного сегмента и его фрагменты. Плоский фронт падает снизу вверх.

Исследования показывают, что использование Янус частиц также способствует увеличению добротности и увеличению составляющих электрической и магнитной напряженности вблизи поверхности удаленного элемента частицы. Эффект является резонансным по отношению к объему удаляемой фракции вещества и наблюдается в диапазоне параметра размеров q=30…90.

В тоже время, введение микросфер в обычную систему оптического микроскопа приводит к нежелательным аберрациям, которые снижают контрастность и качество изображения, несмотря на то, что наблюдается улучшение локального разрешения в области под микросферой. В тоже время в [116] впервые было продемонстрировано, что диэлектрический куб из Тефлона с размером грани, равной длине волны, можно использовать для улучшения разрешения путем простого его размещения в сфокусированной области системы визуализации (линзы), независимо от значения числовой апертуры. Этот принцип может быть использован как в системах «на отражение», так и в системах передачи изображений «на пропускание» [117]. Используя частоту 125 ГГц в системе формирования изображений, мы получили разрешение, соответствующую частоте 275 ГГц, что в 2,2 раза выше. Это особенно важно в ТГц диапазоне, где увеличение частоты зондирующего излучения приводит к увеличению поглощения в среде, и поэтому часто не применимо. При этом коэффициент увеличения контрастности изображения составлял примерно 4,4.

Наименьшее разрешение в микроскопии оптического типа ограничивается фундаментальным пределом дифракции. Микроскопы на основе диэлектрических микрочастиц размера порядка длины волны являются многообещающим инструментом для преодоления дифракционного предела. Однако микрочастицы имеют низкий контраст изображения в воздухе, что ограничивает применение этого метода. Увеличить контраст изображения возможно, используя микрочастицы, обеспечивающих формирование области локализации излучения под углом к направлению падения излучения (под углом к оптической оси), например, фотонный крючок или фотонную струю с наклонным освещением в ближнем поле [118]. По аналогии с классическим оптическим микроскопом [119,120], в случае осевого формирования фотонной струи излучение падает на образец, и дифрагированные порядки -1 и +1 находятся за пределами границы частицы, а наклонное облучение приводит к тому, что дифрагированный +1 порядок попадает на частицу, что в итоге позволяет увеличить контраст изображения [121] – рис.9.

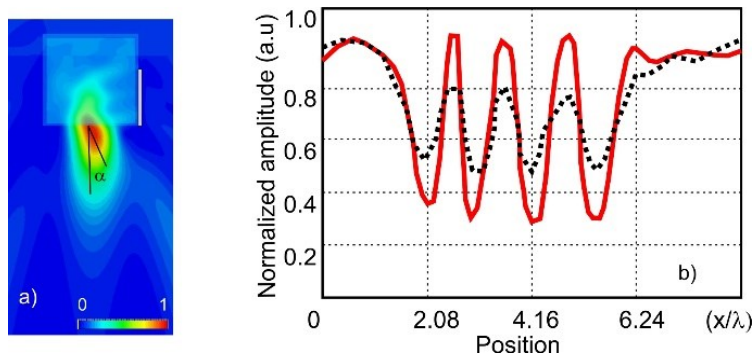

Рис.9. (a) Формирование фотонного крючка для кубической частицы с боковой маской, (b) контраст изображения тестового объекта при облучении фотонной струей (черный) и фотонным крючком (красный).

**Наноструктурированные частицы с показателем около 2**

Как упоминалось выше, поперечный размер области локализации поля диэлектрической мезомасштабной частицы обычно больше λ/3 при показателе преломления менее критического. В [122,123] впервые показано, что размер области локализации электромагнитного поля, которое формируется около теневой поверхности диэлектрической мезомасштабной сферы или кубика с показателем преломления материала около 2, может быть значительно уменьшен путем введения наноотверстий на ее теневой поверхности, которая улучшает пространственное разрешение вплоть до λ/100. Введение наноотверстия на теневой поверхности диэлектрической частицы позволяет «сжать» локализацию поля, характерную для фотонной струи, до размера этого наноотверстия благодаря контрасту диэлектрической проницаемости между частицей и наноотверстием, а минимальная величина перетяжки пучка определяется не длиной волны, а размером этого отверстия.

Такое наноструктурирование теневой поверхности частиц с показателем преломления около 2 позволило предложить концепции оптического пылесоса [123] и оптического магнита [124]) для манипуляции наночастицами, а также оптического модулятора излучения на основе комбинации мезоразмерной частицы и графенового монослоя [125].

Более сложное наноструктурирование теневой поверхности сферических частиц с показателем преломления материала менее 2 на основе дифракционных структур применительно к оптическому манипулированию рассмотрено в [126].

**структурированные поля в виде фотонных крючков и петель**

Недавно было открыто новое семейство ближнепольных локализованных изогнутых световых лучей, известных как «фотонные крючки», которые принципиально отличается от семейства лучей Эйри [127]). Термин «фотонный крючок» и его концепция были введены в [128, 35]. Фотонные крючки (ФК) уникальны в том смысле, что их радиус кривизны значительно меньше длины волны, что означает, что такие пучки могут испытывать максимальное ускорение, а боковые лепестки не повторяют форму основного пучка и не изогнуты. Более того, в ФК существует точка перегиба, при котором пучок меняет направление своего распространения. ФК можно непосредственно наблюдать вблизи теневой поверхности мезомасштабной Янус частицы, и он свободен от каких-либо других ограничений. Этими свойствами не обладают пучки типа Эйри [129].

В простейшем случае Янус частицы в виде диэлектрического куба с призмой [128, 35, 130, 131], интерференция, вызванная разной толщиной частиц вдоль направления поляризации, приводит к появлению дополнительных сингулярных точек вблизи теневой поверхности асимметричной частицы, в частице неравномерно изменяется время полной фазы колебаний оптической волны, что в конечном счете вызывает искривление пучка.

На сегодня известно несколько подходов к формированию искривленных пучков типа фотонный крючок, систематизированных и подробно рассмотренных в [132]. Среди них: несимметрия формы частицы при постоянном показателе преломления [133,134], симметричная частица с несимметрией показателя преломления [135-137], симметричная однородная частица с ассимметрией облучающей волны [138,139], а также симметричные частицы, расположенные в неоднородной (например, с дискретным распределением показателя преломления) среде и комбинация этих методов, использование системы мезоразмерных рассеивателей [140]. Эта концепция может быть также развита для получения ближнепольных пучков типа фотонная петля [35].

Концепция фотонного крючка предлагает изысканный контроль над движением частицы для манипулирования и сортировки клеток на платформах "лаборатория-на-чипе". Уникальное свойство ФК – субволновая кривизна пучка – может быть использовано для транспорта наночастиц за преграду [128]. Так, в приближении Рэлеевкой наночастицы было показано, что возможно ее перемещение вокруг диэлектрической пластины-препятствия [141]. Некоторые из возможных биомедицинских применений этой концепции *in vitro* заключаются в том, чтобы направлять клетки по изогнутой траектории для последующей их дифференциации [142], а также для транспорта нанообъектов в межклеточной или внутриклеточной среде по искривленной траектории, для доставки нанолекарств через барьеры центральной нервной системы [143], лазерной хирургии [144], микрофлюидных устройствах, наноструктурирования поверхностей, новых фотонных устройствах [145] и других областях.

Контроль ближнего поля будет становиться все более и более важным в интегрированной фотонике. В этой связи структурированные пучки типа фотонных струи и крючков вполне могут быть тем мостом, который связывает фотонный и материальный миры вместе. В этой связи интересно также, что свойства мезомасштабных пирамидальных частиц [36,37, 146-148] раскрывают возможную природу египетских пирамид [149,150].

**Низкоразмерные системы. Плазмонные струи и крючки**

Изучение свойств низкоразмерных систем занимает одно из центральных мест в фотонике. Плазмонные структуры интенсивно изучаются из-за способности локализовывать свет на субволновых масштабах [151,152]. Поверхностные плазмонные волны, по сути, это двумерные волны, компоненты поля которых экспоненциально затухают при удалении от границы раздела метал-диэлектрик.

По-видимому, впервые возможность формирования аналога фотонной струи для плазмонных волн на основе диэлектрического диска, расположенного на металлической пленке, была теоретически показана в [153]. Позднее, на основе решения уравнений Максвелла, образование плазмонной струи на основе диэлектрического кубоида из нитрида кремния (Si3N4) было исследовано в [154]. При этом, используя идею волновода на основе цепочек кубических частиц [87], в [155]) была продемонстрирована возможность увеличения длины распространения локализованной плазмонной волны. Однако экспериментально эффект плазмонной струи на основе диэлектрического кубика, расположенного на золотой пленке, был впервые подтвержден в [156]. Эффект плазмонной струи продемонстрировал захватывающие потенциальные применения в интегрированной оптике и оптике ближнего поля для управления плазмонными волнами. Такая новая и простая платформа может обеспечить новый путь для плазмоники, визуализации с высоким разрешением, биофотоники, хранения оптических данных и будущих приложений от управления светом в наномасштабе до фотонных устройств на кристалле [157].

Это же относится и к плазмонному крючку. Теоретически эффект плазмонного крючка был обоснован в [158]. И совсем недавно он был впервые подтвержден экспериментально [159]. Важно отметить, что плазмонный крючок продемонстрировал *наименьшую* кривизну луча, когда-либо зарегистрированную для плазмонных волн. Экспериментальная проверка этих эффектов открывает новые возможности для управления взаимодействием между светом и веществом на наноуровне и могут значительно расширить доступные инструменты для новых интригующих манипуляций со светом [160].

**Акустические струи и крючки**

В оптическом диапазоне для характеризации фотонных струй обычно достаточно двух параметров: параметра размера Ми и контраста показателя преломления. В терагерцовом и СВЧ диапазонов добавляется параметр, характеризующий потери в материале частицы. В акустическом диапазоне добавляется еще один параметр - все материалы, как правило, являются анизотропными ввиду наличия двух скоростей звука в материале твердой частицы: продольной и поперечной (shear wave).

Термин «акустическая струя» был введен в работе [161] на основе аналогии уравнений Максвелла и механики сплошных сред для линейного режима. В [162,163] было экспериментально продемонстрировано получение акустической струи с минимальным размером перетяжки пучка акустической струи около половины длины волны для сферической и цилиндрической частиц. Было показано существенное влияние на формирование акустической струи соотношения продольной и поперечной скоростей звука в материале частицы [164]. Исследованы акустические струи на основе частиц из метаматериалов [161,165]. Был также подтвержден эффект аномальной аподизации для акустических струй как в газе, так и жидкости [166,167].

В результате исследований было показано, что за условия формирования акустической струи, также как и в оптике, существенная роль отводится системе акустических вихрей как в материале частицы, так и в области ее теневой поверхности.

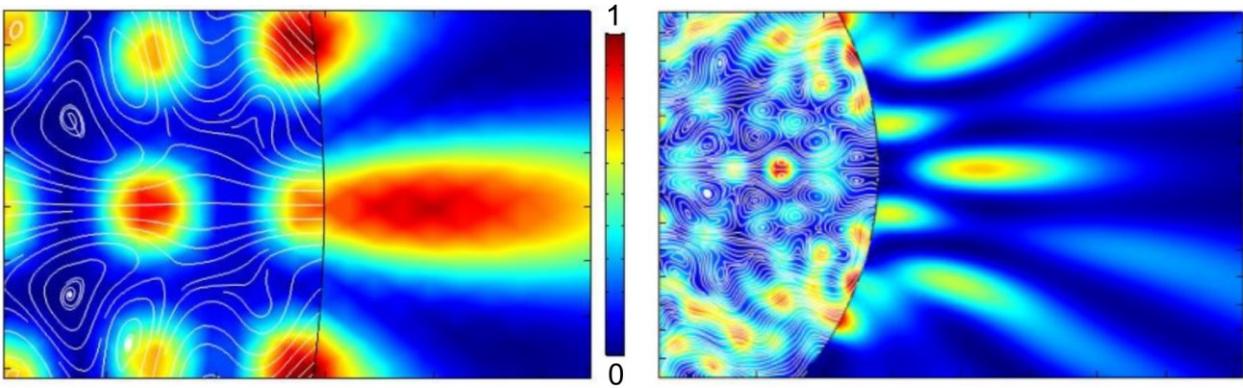

Рис.10. Вихревые течения в материале частицы (Рексолит, плотность 1049 kg/$m^3$, продольная скорость звука $C_l$=2337 m/s, поперечная скорость звука $s$=1157 m/s)) и образование горячих точек в нерезонансном (слева) и вблизи резонанса (справа) режимах.

Кроме того, оптический эффект суперрезонанса наблюдается и в акустике [168,169] – рис.11. Вблизи резонансной частоты разрешение в области горячей точки составляет около 0.21λ - 0.23λ.

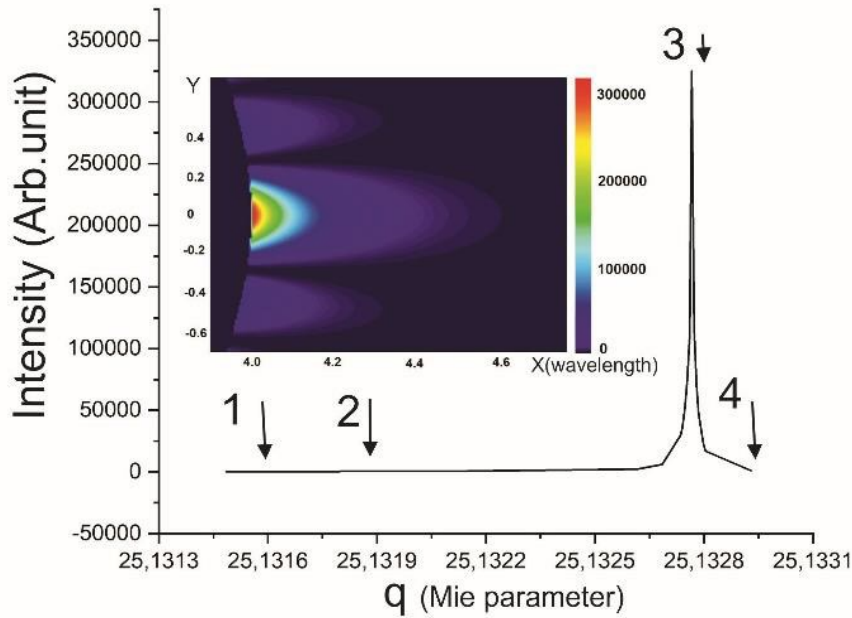

Рис.11. Резонансное рассеяние акустической волны на сферической частице из Рексолита (Rexolite), расположенной в воде, в зависимости от параметра размера Ми. На вставке: структура «горячей точки» на теневой поверхности частицы на резонансной частоте. Максимальная интенсивность в указанных точках 1-4 составляет: 325 (1), 2500 (2), 325000 (3), 1000 (4) в единицах интенсивности падающей на частицу акустической волны.

Концепция фотонного крючка была распространена на акустические волны и экспериментально подтверждена в [170]. Этот изогнутый акустический пучок генерировался с помощью погруженной в воду частицы из Рексолита с трапециевидной прямоугольной призмой, на которую падала плоская ультразвуковая волна с частотой 250 кГц. Аналогичные результаты были получены позднее в [171]. В [172] акустические струи и акустические крючки в воде были получены с помощью мезомасштабного (параметр q<30) цилиндра из АБС-пластика, изготовленного на 3D-принтере. Было продемонстрировано, что в резонансном режиме возбуждение мод шепчущей галереи приводит к минимальным значениям диаметра акустической струи менее дифракционного предела. В нерезонансном режиме внутри цилиндра размещался кубический фононный кристалл, и в зависимости от его ориентации относительно направления падения волны формировались как акустические струи, так и акустические крючки. Позднее, частица, состоящая из полусферического цилиндра из Рексолита и полусферического фотонного кристалла (использовалось два разных материала с разными эффективными показателями преломления), формирующая акустический крючок, была рассмотрена в [173].

Структурированные акустические пучки типа струи и крючков могут также найти потенциальное применение в микроскопии, медицинской визуализации [174], мониторинге состояния различных конструкций [175], в сборе и локализации энергии [176].

В целом, наличие также целого ряда новых приложений свидетельствует о том, что в оптике, акустике и плазмонике возникло новое перспективное направление.

## Список литературы